\documentclass[a4paper,11pt]{article}
\usepackage[utf8]{inputenc}
\usepackage{physics}
\usepackage{amsmath}
\usepackage{multirow}
\usepackage{pos}
\usepackage{siunitx}
\usepackage{simpler-wick}
\usepackage{xcolor}

\newlength{\ftsize}

\newcommand{\scalecomment}{We use the reference value $\sqrt{8t_0} = 0.415\;\text{fm}$ \cite{Bruno2017} to obtain physical units. }

\title{Hadronic vacuum polarization with C* boundary conditions}

\author*[a]{Anian Altherr}
\author*[a]{Roman Gruber}
\author[b]{Lucius Bushnaq}
\author[c]{Isabel Campos}
\author[a]{Marco Catillo}
\author[d]{Alessandro Cotellucci}
\author[d,e,f,g]{Madeleine Dale}
\author[b]{Patrick Fritzsch}
\author[a]{Javad Komijani}
\author[d,h]{Jens Lücke}
\author[a]{Marina Krsti\'{c} Marinkovi\'{c}}
\author[i]{Sofie Martins}
\author[d,h]{Agostino Patella}
\author[e,f]{Nazario Tantalo}
\author[a]{Paola Tavella}

\affiliation[a]{Institut für Theoretische Physik, ETH Zürich, Zürich, Switzerland }
\affiliation[b]{School of Mathematics, Trinity College Dublin, Dublin, Ireland }
\affiliation[c]{Instituto de Física de Cantabria \& IFCA-CSIC, Santander, Spain }
\affiliation[d]{Institut für Physik {\&} IRIS Adlershof, Humboldt Universität zu Berlin, Berlin, Germany }
\affiliation[e]{Università degli Studi di Roma "Tor Vergata", Rome, Italy}
\affiliation[f]{INFN, Sezione di Tor Vergata, Rome, Italy}
\affiliation[g]{Department of Physics, University of Cyprus, Nicosia, Cyprus}
\affiliation[h]{DESY, Zeuthen, Germany }
\affiliation[i]{University of Southern Denmark, Odense, Denmark}

\emailAdd{aaltherr@ethz.ch}
\emailAdd{rgruber@ethz.ch}

\date{August 2022}
\abstract{

    We present a progress report
    on the calculation of the connected hadronic
    contribution to the muon $g-2$
    with C$^\star$ boundary conditions.
    For that purpose we use
    a QCD gauge ensemble with 3+1 flavors and two QCD+QED gauge ensembles with 1+2+1 flavors of dynamical quarks generated by the RC$^\star$ collaboration. We detail the calculation
    of the vector mass and elaborate on both statistical and systematic
    errors.
}

\FullConference{%
The 39th International Symposium on Lattice Field Theory,\\
8th-13th August, 2022,\\
Rheinische Friedrich-Wilhelms-Universität Bonn, Bonn, Germany
}

\begin{document}

\maketitle

\section{Introduction}
There has been a tantalizing discrepancy between the experimental and theoretical
values for the anomalous magnetic moment of the muon \cite{Aoyama2020}. On the experimental side,
two experiments at Fermilab \cite{Abi2021} and J-PARC \cite{Abe2019} are aiming to further decrease 
the uncertainty in the next few years. Hence, progress on the theoretical
side is crucial to keep up with experiments. The main uncertainty
in the theoretical calculation stems from the hadronic vacuum polarization 
(HVP) contribution.
To keep up with experimental precision, a subpercent precision is required
for the HVP contribution. While there exist several sub-percent
calculations of the connected light contribution 
(see Ref.~\cite{Aoyama2020} for an overview) at the 
isospin-symmetric point, the systematic uncertainty due to
neglecting isospin-breaking effects becomes crucial. Several collaborations
have started to take these effects into account for the
anomalous magnetic moment 
\cite{Borsanyi2021,Giusti2019,Chakraborty2018,Milc2020,Blum2018,Lehner2020,Risch2021}.

There exist different ways of including these effects; for a detailed
review of some of these methods see Ref.~\cite{Patella2016,Tantalo2022}. Among these is
$\text{QED}_\mathrm{L}$ that removes the zero modes of the photonic field by hand
\cite{Uno2008} and thus creates a non-local field theory. 
Alternatively, $\text{QED}_\mathrm{m}$
uses a finite photon mass as an infrared regulator \cite{Endres2015, Clark2022},
but requires an extrapolation to zero photon mass.
In $\text{QED}_{\infty}$ the QED part is analytically calculated in the continuum \cite{feng2019,Christ2021,Blum2017}.
In this proceedings, we choose $\text{QED}_\mathrm{C}$ 
in which the zero modes of the photon field are absent
due to C$^\star$ boundary conditions
\cite{Kronfeld1991, Wiese1992,Kronfeld1993,Lucini2016,Hansen2018,polley1993},
thus allowing for a local field theory without including an additional
regulator. Using these boundary conditions,
we explore the connected HVP contribution.

After a brief introduction to the HVP in Section~\ref{sec:hvp}, we introduce
a few aspects of C$^\star$ boundary conditions in Section~\ref{sec:cstar}.
 In Section~
\ref{sec:results} we present our results of vector masses and
the HVP and conclude in Section~\ref{sec:conclusion}.

\section{Hadronic vacuum polarization}
\label{sec:hvp}

The HVP to the muon anomalous magnetic moment can be obtained
using the time-momentum representation \cite{bernecker2011}:
\begin{align} \label{eq:a_mu}
    a_{\mu}^{\text{HVP}} = \left(\frac{\alpha}{\pi} \right)^2
            \int_0^{\infty} \mathrm{d}x_0
            G(x_0) \tilde{K}(x_0; m_\mu).
\end{align}
Here $\alpha$ is the electromagnetic coupling, $\tilde{K}(x_0; m_\mu)$ is a kernel
function defined in Ref.~\cite[Eq.~(8)]{DellaMorte2017} with the muon mass $m_\mu$, and $G(x_0)$
is expressed in terms of the two-point correlation function of the electromagnetic
current $j_k(x)$:
\begin{equation} \label{eq:twopoint}
    G(x_0) = - \frac{1}{3} \sum_{k=1}^3 \int \mathrm{d}^3x \ev{j_{k}(x) j_{k}(0)}.
\end{equation}
The integral over time $x_0$ in Eq.~\eqref{eq:a_mu} 
can be split into two parts to take into account the 
finite-time extent on the lattice and 
to treat the noise that dominates the signal for large times separately
\cite{bernecker2011},
see Section~\ref{sec:mass}.

\section{C\texorpdfstring{$^\star$}{*} boundary conditions}
\label{sec:cstar}

At the target precision of one percent it is
crucial to include QED effects in lattice simulations, as they are expected to be of the order of one percent.
A naive implementation of QED on a finite-volume lattice with periodic boundary conditions does not allow to simulate charged particles because states with non-zero electric charge violate Gauss' law.
 C$^\star$ boundary conditions \cite{Kronfeld1991,Kronfeld1993,Wiese1992,Lucini2016,Hansen2018} provide a remedy, 
and they do not lead to a non-local field theory as opposed to $\mathrm{QED}_\mathrm{L}$.
The fields obey the following constraints in spatial direction $\hat{k} = \hat{1}, \hat{2}, \hat{3}$

\begin{equation}
\begin{aligned} \label{eq:orbi}
    \psi_f(x + L \hat{k}) &= \psi_f^c(x) := C^{-1} \overline{\psi}_f^T(x)
    \\
    \overline{\psi}_f(x + L \hat{k}) &= \overline{\psi}^c(x) := - \psi_f^T(x) C
    \\
    U_{\mu}(x + L \hat{k}) &= U_{\mu}^c(x) := U_{\mu}(x)^{\ast}
    \\
    A_{\mu}(x + L \hat{k}) &= A_{\mu}^c(x) := -A_{\mu}(x),
\end{aligned}
\end{equation}
where $\psi_f$ and $\overline{\psi}_f$ are the fermionic fields of flavor $f$,
$U_{\mu}(x) \in SU(3)$ are the QCD gauge links and $e^{iA_{\mu}(x)} \in U(1)$ are the QED gauge links; $U_\mu^*$ denotes complex conjugation and
we denote charge conjugation by the superscript $c$. The charge 
conjugation matrix $C$ obeys 
    $C \gamma_{\mu} C^{-1} = - \gamma_{\mu}^T$
with the Euclidean gamma matrices $\gamma_{\mu}$.
We note that the photon field $A_{\mu}(x)$ is antiperiodic and
thus does not have a zero-momentum component by construction.

\subsection{Correlation functions}
It is useful to combine $\psi_f$ and $\overline{\psi}_f$ into a spinor doublet
\begin{equation} \label{eq:doublet}
    \chi_f(x) = \begin{pmatrix}
        \psi_f(x) \\
        \psi_f^c(x)
    \end{pmatrix}
\end{equation}
\begin{sloppypar}
\noindent and express the action in terms of $\chi_f$. When using the Wilson-Dirac
formulation with a Sheikholeslami-Wohlert improvement term
(see Refs.~\cite{DiracPatella,CstarPatella} for details), we find that the 
fermionic action can be written as
\end{sloppypar}
\begin{equation}
    S = - \frac{1}{2} \sum_{f=1}^{N_f}\chi_f^T C \sigma_1 D \chi_f,
\end{equation}
where $\sigma_1$ is the first Pauli matrix
acting on the spinor doublet in Eq.~\eqref{eq:doublet} and $N_f$ is the
number of flavors. We note that
$D$ has $24V \times 24 V$ complex components as it acts on a spinor doublet ($V$ denotes
the lattice volume).
Wick contractions yield (see Ref.~\cite[Appendix B]{Hansen2018})
\begin{equation}
    \wick{ \c\chi_f(x) \c\chi_{f'}(y)^T} = - \delta_{f,f'} D_{f'}^{-1}(x; y) \sigma_1 C^{-1}.
\end{equation}
To evaluate the two-point function in Eq.~\eqref{eq:a_mu} we 
use
\begin{equation} \label{eq:jmu}
    j_\mu(x) = \frac{1}{2} \sum_{f=1}^{N_f} q_f \chi_f^T(x) \sigma_3 \sigma_1 C \gamma_\mu \chi_f(x),
\end{equation}
where
$q_f$ the charge of flavor $f$,
$\sigma_3$ the third Pauli matrix acting on the spinor doublet.
When expressed in terms of $\psi_f$ and $\overline{\psi}_f$
, $j_\mu(x)$ is equivalent to the familiar expression
$\sum_{f} q_f \overline{\psi}_f(x) \gamma_\mu \psi_f(x)$.
Inserting Eq.~\eqref{eq:jmu} into the two-point correlation function, we find
\begin{align*}
    \ev{j_\mu(x) j_\nu(y)} =
    \ev{j_\mu(x) j_\nu(y)}_{\text{conn}} +
    \ev{j_\mu(x) j_\nu(y)}_{\text{disc}},
\end{align*}
where in this proceedings we only take the connected contractions into account\footnote{We note that
there are two combinations of Wick contractions that contribute to the connected part.}:
\begin{equation}
    \ev{j_\mu(x) j_\nu(y)}_{\text{conn}}
    = \frac{1}{2} \sum_{f=1}^{N_f} q_f^2
        \tr[ \sigma_3 D_f^{-1}(x; y) \gamma_\nu \sigma_3 D_f^{-1}(y; x) \gamma_\mu ],
\end{equation}
and neglect the disconnected contribution
\begin{equation}
    \ev{j_\mu(x) j_\nu(y)}_{\text{disc}}
    = \left(\frac{1}{2} \sum_{f=1}^{N_f} q_f
        \tr\left[ \sigma_3 \gamma_{\mu} D_f^{-1}(x; x) \right] \right)
        \left(\frac{1}{2} \sum_{f=1}^{N_f} q_f \tr\left[ \sigma_3 \gamma_{\nu} D_f^{-1}(y; y) \right]\right).
\end{equation}
This expression is similar to the one obtained in periodic boundary 
conditions, the only difference being that $D^{-1}$ has $24V \times 24V$
components.

\subsection{Lattice parameters}
We use configurations generated by the RC$^\star$ collaboration \cite{Bushnaq2022}
with 3+1 and 1+2+1 flavors of Wilson quarks with a clover term for 
both $SU(3)$ and $U(1)$. The $SU(3)$ action
is $\mathcal{O}(a)$-improved in a non-perturbative way for QCD ensembles \cite{Fritzsch2018},
 which is valid for QCD+QED ensembles up to $\mathcal{O}(\alpha)$, and the $U(1)$ action is (so far) tree-level improved.
All configurations use
C$^\star$ boundary conditions (see Section~\ref{sec:cstar}) and are summarized in 
Table~\ref{tab:ensembles}. There is one QCD ensemble (\texttt{A400a00b324})
and two ensembles with dynamically generated QCD+QED fields
(\texttt{A360a50b324} and \texttt{A380a07b324}) at two values
of the fine structure constant; one is close to the physical value and one is larger.
The charged pion mass ranges between
approximately $360~\mathrm{MeV}$ and $400~\mathrm{MeV}$
and we have $m_{\pi^{\pm}} L \approx 2.9 - 3.5$.
\scalecomment
We refer to Ref.~\cite{Bushnaq2022} for the details of the generation.

\begin{table}
    \centering

    \begin{tabular}{r | r r r}
     ensemble                      & A360a50b324          & A380a07b324          & A400a00b324  \\ [0.5 ex]
     \hline
     flavors                       & 1 + 2 + 1            & 1 + 2 + 1                   & 3 + 1       \\ 
     $\beta$                       & 3.24                 & 3.24                        & 3.24        \\ 
     $\alpha_R$                      & $0.040633(80)$         & $0.007081(19)$  & $0.0$       \\ 
     $m_{\pi^{\pm}} [\mathrm{MeV}]$      & $358.6(3.7)$             & $383.6(4.4)$                      & $398.5(4.7)$    \\ 
     $a [\mathrm{fm}]$             & 0.05054(27)          & 0.05323(28)                 & 0.05393(24) \\ 
     number of used configurations                       & 181                  & 200                         & 200         \\ 
    \end{tabular}

    \caption{Summary of used ensembles and their parameters, compare Ref.~\cite{Bushnaq2022} for details. All lattices
    have size $32^3 \times 64$. Ensembles with 1+2+1 flavor decomposition have degenerate
    down and strange quark, ensembles with 3+1 flavor decomposition have degenerate up, down and strange quark. 
    The name of the ensemble --- say \texttt{A360a50b324} --- contains the approximate charged pion mass 
    ($360~\mathrm{MeV}$), the bare fine-structure constant ($0.050$) and the coupling ($3.24$).
    We denote the renormalized fine-structure constant by $\alpha_R$ and also tabulate the number of used configurations.
    Note that we have not used all available configurations, which is about 2000 for each ensemble.
    \scalecomment}
    \label{tab:ensembles}
\end{table}

\section{Preliminary results}
\label{sec:results}
    In this section, we first examine the signal-to-noise ratio of the two-point function and then
    show values for the vector masses and the HVP.
    Our analysis is based on the openQ*D program package \cite{Campos2020}. 

\subsection{Signal-to-noise}
\begin{figure}
        \centering
        \includegraphics[width=\textwidth]{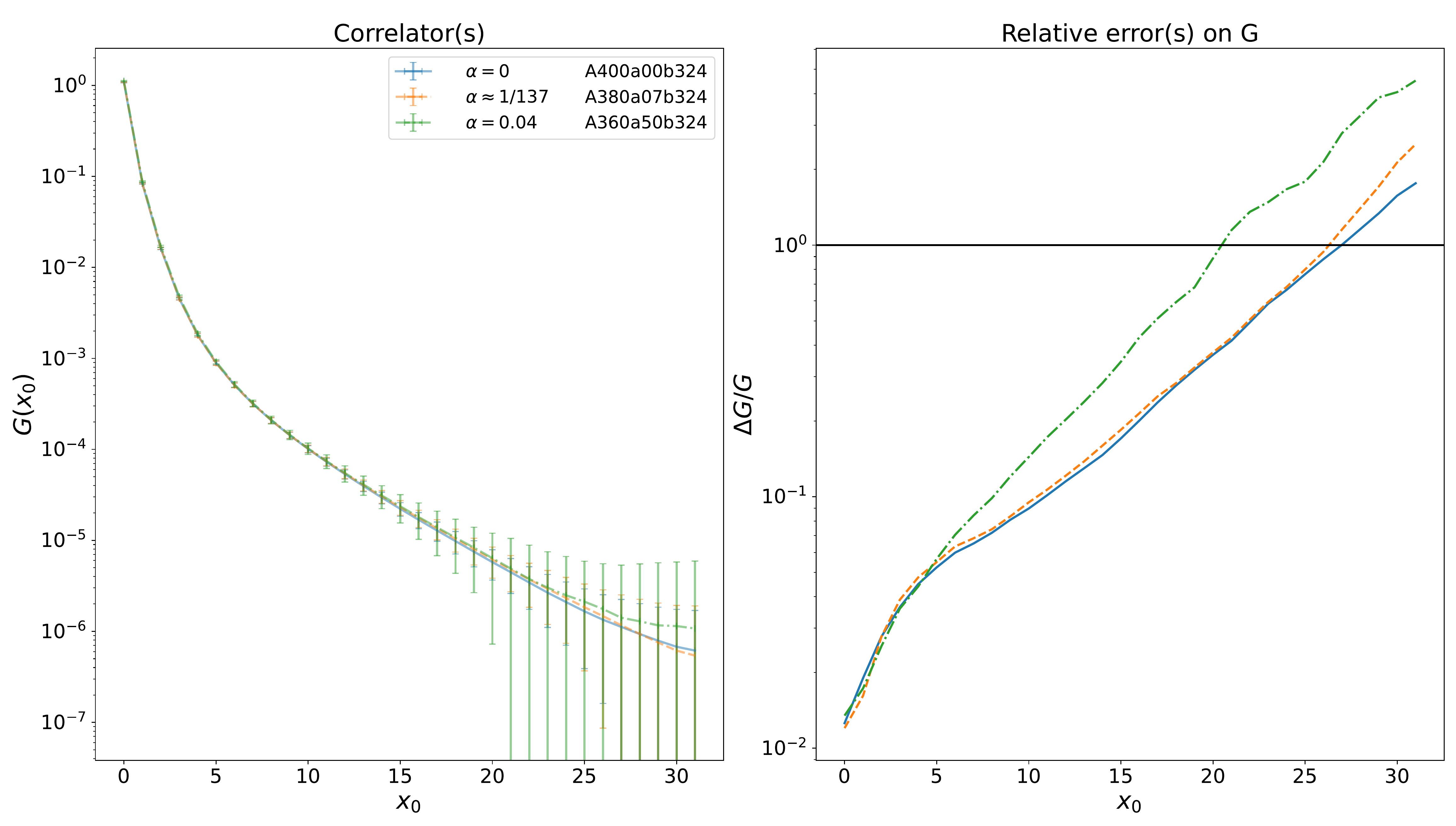}   
        \caption{Relative error comparison. On the left-hand side, we plot the local-local two-point function of the electromagnetic current
        $G(x_0)$, on the right-hand
        side we plot the relative errors. The errors for the QCD ensemble (blue solid line) and the QCD+QED ensemble with physical
        $\alpha$ have comparable relative errors, whereas the relative error for the QCD+QED ensemble with larger $\alpha$ is 
        slightly larger. The number of used gauge configurations and stochastic sources are $181$ and $10$, respectively in all three cases.}
        \label{fig:noise}
    \end{figure}

    Using the three ensembles in Table~\ref{tab:ensembles} we calculate the
    two-point function in time-momentum representation $G(x_0)$.
    As illustrated in Figure~\ref{fig:noise}, the relative statistical error for simulating QCD (\texttt{A400a00b324}) is comparable to the error of simulating QCD+QED at physical $\alpha$ (\texttt{A380a07b324}).
    On the other hand, simulating QCD+QED at unphysically large $\alpha$ (\texttt{A360a50b324}) yields larger errors than for physical $\alpha$.
    This might be due to a lower pion mass ($360~\mathrm{MeV}$ for \texttt{A360a50b324}) compared to the other ensembles.

    \subsection{Mass spectroscopy}

\label{sec:mass}

    As the signal-to-noise ratio deteriorates for large times $x_0$,
    we introduce a $x_{0,\text{cut}}$ and replace the two-point function $G
    (x_0)$ by a model function for $x_0 > x_{0,\text{cut}}$.
    In finite volume the two-point function $G(x_0)$ can be written as a sum of exponentials with positive coefficients. 
    We choose to only consider the leading term in the spectral decomposition
    for our model function
        \begin{align} \label{eq:model}
            G(x_0)_{\text{model}} = A e^{- m_{\text{eff}} x_0},
        \end{align}
    where $m_{\text{eff}}$ is the effective mass of the corresponding vector meson ground state in lattice units and $A$ the decay amplitude.

    In order to determine the coefficients $m_{\text{eff}}$ and $A$, we perform
    a $\chi^2$-fit to the two-point correlation function in a fit range
    $I_{\text{fit}}$ where excited states have decayed sufficiently, but
    where there is still a clear signal.
    Since we are interested in the ground state, we apply Gaussian smearing
    to the sink and source point to increase the overlap with the ground
    state. In addition, we apply stout smearing to
    the gauge fields. 
    The effective mass $m_{\text{eff}}$ is then extracted by using a 
    $1$-parameter logarithmic fit to the correlator on a subset 
    of $40$ configurations. 
    Finally, we use the effective mass as input for the $1$-parameter fit
    of the decay amplitude
    to the two-point function (obtained with point sources, without
    smearing).
    In the future we plan to improve the model function by 
    including excited states.



Table \ref{tab:masses} shows a compilation of the extracted masses. The statistical error is obtained using the jackknife method and the systematic error due to choosing a fit range for the vector mass is estimated by varying the range $I_{\text{fit}}$. The total error in the table equals the two error contributions added in quadrature. The mass of the ground state for the charm contribution is not determined because the model part of the correlator has a negligible contribution to the HVP in that case (see Figure \ref{fig:integrand}).

\begin{table}
    \centering

    \begin{tabular}{llll}
     ensemble & quark & $m_{\text{eff}} $ & $\chi_r^2$ \\
     \hline
     \multirow{2}{*}{A360a50b324} & down/strange & 0.262(7) & 1.03 \\
        & up & 0.267(8) & 0.93 \\
     \hline
     \multirow{2}{*}{A380a07b324} & down/strange & 0.265(6) & 1.07 \\
        & up & 0.266(4) & 0.95 \\
    \hline
     A400a00b324 & up/down/strange & 0.278(7) & 0.94 \\
    \end{tabular}

    \caption{Vector meson mass for the three ensembles: The effective mass
    $m_{\text{eff}}$ is obtained by performing a $\chi^2$-fit of the
    two-point function Eq.~\eqref{eq:twopoint} to a single exponential
    Eq.~\eqref{eq:model}. In the last column, we display the $\chi^2$
    value per degree of freedom $\chi_r^2$.
    }
    \label{tab:masses}
\end{table}

    \subsection{Hadronic vacuum polarization contribution}
\begin{figure}
        \centering
        \includegraphics[width=0.7\textwidth]{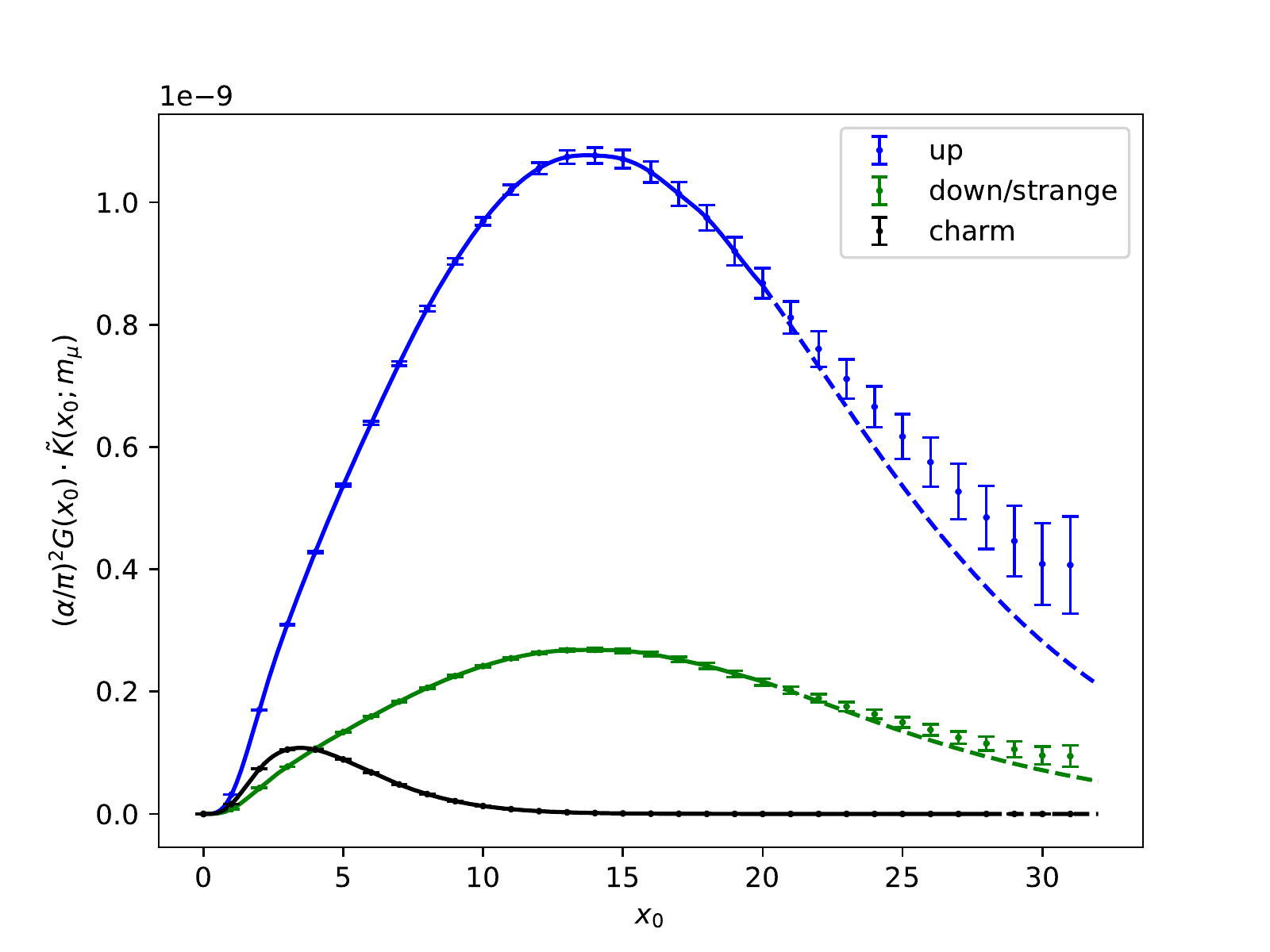}
        \caption{Integrand of the HVP contribution for the \texttt{A380a07b324} ensemble
        where we use a conserved-local two-point function. Points label
        the actual lattice data, solid lines symbolize interpolation (for $x_0 < x_{0, \text{cut}}$) and dashed
        lines stand for the model function ($x_0 \geq x_{0, \text{cut}}$)
        respectively. The systematic error occurring due to using only a single exponential for the model part is significant for 
        the up contribution (blue) and down/strange contribution (green); for the charm contribution (black) it is negligible.}
        \label{fig:integrand}
    \end{figure}

    As a preliminary check we plot the integrand of Eq.~\eqref{eq:a_mu}
    in Figure~\ref{fig:integrand}.
    Here we use a combination
    of conserved and local currents, i.e., $G(x_0)$ is obtained from
    $\ev{ j_k^{\text{conserved}}(x) j_k^{\text{local}}(0)}$.
    The use of local currents simplifies the calculations, but requires
    both a multiplicative as well as an additive renormalization constant
    \cite{Collins2006}. We set the renormalization constant due to
    the local current to unity. 
    The results for the HVP contribution of
    muon $g-2$ are shown in Table~\ref{tab:amu}. 

    \begin{table}
    \centering
    \begin{tabular}{llS}
    ensemble & flavor & \multicolumn{1}{c}{$a_\mu^{\mathrm{HVP}} \times 10^{10}$} \\
    \hline  
    \multirow{3}{*}{A360a50b324} & up &                  309(11) \\
     & down/strange &                                            77(2) \\
      & charm &                                                  10.62(11) \\
      \hline
    \multirow{3}{*}{A380a07b324} & up &        331(7) \\
     & down/strange &                                            83(2) \\
     & charm &                                                   9.78(10) \\  
     \hline
    \multirow{2}{*}{A400a00b324} & up/down/strange &     319(8) \\
     & charm &                                                  9.97(9) \\
    \end{tabular}
    \caption{Results for the HVP contribution. The ensembles employ
    C$^\star$ boundary conditions in spatial directions, have a pion mass
    of approximately 360, 380, and 400 MeV respectively, and 
    are simulated on a lattice with extent $32^3 \times 64$ (compare
    Table~\ref{tab:ensembles}). We use a combination of conserved and local current and 
    set the renormalization constant to unity.}
    \label{tab:amu}
    \end{table}

\subsection{Error contributions}
    We display the error contributions for the up-quark contribution of the QCD+QED ensemble with physical $\alpha$
    in Table~\ref{tab:errors}. 
    An estimate of the statistical Monte Carlo error is obtained
    using jackknife.
    The error occurring in the determination of the vector mass (compare Table~\ref{tab:masses})
    affects the precision of the HVP contribution as does the error on the lattice spacing.
    The lattice spacing has been determined in Ref.~\cite{Bushnaq2022}, which is based on the value of the the gradient flow
    scale $t_0$ determined in Ref.~\cite{Bruno2017};
    the relative error of the lattice spacing turns out to be
    $\frac{\Delta a}{a} = 0.53 \%$ (see Table~\ref{tab:ensembles}).
    We note that the gradient flow scale has only been determined for QCD ensembles,
    a scale setting in full QCD+QED
    remains to be worked out; see Ref.~\cite{Tantalo2022}.
    The contribution to the uncertainty on the HVP due to error propagation
    of the uncertainty in scale setting is expected to
    be roughly $1.8 \frac{\Delta a}{a}$ (see Ref.~\cite[Appendix B.2]{DellaMorte2017}), which agrees with our findings.
    In addition to statistical errors, we have systematic errors.
    Due to the exploratory nature of this proceedings
    we do not carry out a continuum extrapolation nor a chiral
    extrapolation to physical meson masses. Neither do we assess
    finite-size effects. There are further systematic errors
    due to the model part. To determine the vector mass and amplitude in 
    Eq.~\eqref{eq:model} we select a fit range and a cutoff $x_{0, \text{cut}}$. We estimate the error due to that choice by varying 
    the cutoff and the fit range around our chosen value. These errors are small compared to the error that arises due to 
    neglecting excited states.
    We estimate the contribution of excited states by
    bounding the correlator from above and estimating the error as the difference
    between the bounds \cite{Borsanyi2021}.
    In the future we plan to reduce the error contributions individually, see Section~\ref{sec:conclusion}.

    \begin{table}
    \begin{center}
    \begin{tabular}{llS}
    & variation w.r.t. & \multicolumn{1}{c}{relative error} \\    
    \hline
    \multirow{3}{*}{statistical} & jackknife &          1.21 \% \\
        & err. prop. of vector mass   &                               1.36 \% \\
     & err. prop. of scale setting &  0.92 \% \\
    \hline
    \multirow{3}{*}{systematic} & fit range &           0.14 \% \\
    & model cutoff $x_{0,\text{cut}}$ &                   0.03 \% \\
    & excited states &                                  1.20 \% \\
    \hline
    total & &                                           2.37 \%
    \end{tabular}
    \end{center}
    \caption{Error contributions of the up contribution for ensemble \texttt{A380a007b324}. The row 'jackknife' quantifies the error due
    to statistical fluctuations in the Monte Carlo simulation. The second row quantifies the error propagation of the uncertainty
    on the vector mass (see Section~\ref{sec:mass}) and the third row the error propagation of the uncertainty on the lattice spacing.
    The error due to selecting a fit range (for the vector mass and the amplitude), and a model cutoff $x_{0,\text{cut}}$ are quantified by varying
    these parameters, the error due to neglecting excited states is estimated by using a
    bounding method. We do not assess continuum extrapolation nor chiral extrapolation, neither do we estimate finite-size effects.} 
    \label{tab:errors}
    \end{table}

\section{Conclusion and outlook}
\label{sec:conclusion}

    We have presented the first calculation of the connected HVP contribution to the muon $g-2$ using
    C$^\star$~boundary conditions. Three different ensembles are used, a QCD ensemble and 
    two QCD+QED ensembles with different values of the fine structure constant. The noise level
    for the ensemble with physical $\alpha$ is comparable to the ensemble with QCD only, whereas for larger $\alpha$
    the noise level increases. There remain a couple of open questions that we 
    plan to address in the future.

    First, isospin-breaking effects have not been addressed in the QCD
    ensemble \verb\A400a00b324\.
    There are two methods to deal with isospin-breaking effects --- a perturbative or a stochastic approach. In the perturbative approach, we expand the QCD+QED action around
    the isospin-symmetric point ($\alpha = 0$, $m_u - m_d = 0$) in $\alpha$ and $m_u - m_d$ and evaluate the correlation function obtained
    in that way \cite{Divitiis2012, Boyle2017}. On the other hand, QED fields can be added stochastically by multiplying QCD gauge links 
    with random $U(1)$-phases 
    \cite{Duncan1996, Boyle2017}. We plan to investigate both approaches.

    Second, finite-size effects for the HVP can be quantified using the Hansen-Patella method \cite{Hansen2020}, where different orders
    of finite-volume effects are generated by powers of 
    $e^{-m_\pi L}$
    with $m_\pi$ the pion mass, and $L$ the linear lattice extent.
    For periodic boundary conditions
    the leading-order term is $\mathcal{O}(e^{-m_\pi L})$ whereas for C$^\star$ boundary conditions this leading order vanishes
    and finite-size effects are of the order $\mathcal{O}(e^{-\sqrt{2} m_\pi L})$ \cite{Martins2021}. In addition, there are power-law
    finite-size effects due to QED, which are expected to be smaller
    compared to periodic boundary conditions (see for example
    Refs.~\cite{Lucini2016} for corrections to baryon masses).

    Third, as seen in Table~\ref{tab:errors}, we need to implement 
    additional variance-reduction methods to achieve a sub-percent
    precision. Firstly, we plan to increase the number of configurations and number of sources to decrease the statistical error and
    to increase the precision of the vector mass determination.
    In addition, low-mode averaging \cite{degrand2004, Bali2010} exploits the structure of the Dirac operator and reduces the variance that stems from
    the low modes. Regarding ensemble generation, multi-level Monte Carlo \cite{DallaBrida2021} reduces the variance in the correlators
    exponentially in the distance. We plan to consider more sophisticated model functions than single exponentials; this will further
    decrease the error due to the model part.

\acknowledgments

We acknowledge access to Piz Daint at the Swiss National Supercomputing Centre, Switzerland under the ETHZ's share with the project IDs s1101, eth8 and go22.
AA's and RG's research is funded by Schweizerischer Nationalfonds Project No. 200021\_200866.
SM received funding from the European Union's Horizon 2020 research and innovation programme under the Marie Sk\l odowska-Curie grant agreement \textnumero~813942.
AC’s and JL’s research is funded by the Deutsche Forschungsgemeinschaft Project No. 417533893/ GRK-2575 “Rethinking Quantum
Field Theory”.

\bibliography{bibliography}
\bibliographystyle{JHEP} 

\end{document}